\documentclass[final,5p,times]{elsarticle}

\usepackage{amsfonts}
\usepackage{eurosym}
\usepackage{amssymb,bbold}
\usepackage{amsmath}
\usepackage{natbib}
\usepackage{hyperref}

\biboptions{sort&compress}

\journal{Physics Letters A}

\hypersetup{
  colorlinks=true,linkcolor=blue,citecolor=blue,
  filecolor=blue,urlcolor=blue,breaklinks=true
}

\begin{document}

\begin{frontmatter}

\title{Exact Green's function for rectangular potentials and
  its application to quasi-bound states}

\author[uepg]{Fabiano M. Andrade}
\ead{fmandrade@uepg.br}

\address[uepg]{
  Departamento de Matem\'{a}tica e Estat\'{i}stica,
  Universidade Estadual de Ponta Grossa,
  84030-900 Ponta Grossa-PR, Brazil
}

\date{\today}

\begin{abstract}
In this work we calculate the exact Green's function for
arbitrary rectangular potentials.
Specifically we focus on Green's function for rectangular
quantum wells enlarging the knowledge of exact solutions for
Green's functions and also generalizing and resuming results in
the literature.
The exact formula has the form of a sum over paths and always
can be cast into a closed analytic expression.
From the poles and residues of the Green's function the bound
states eigenenergies and eigenfunctions with the correct
normalization constant are obtained.
In order to show the versatility of the method, an application of the
Green's function approach to extract information of quasi-bound states
in rectangular barriers, where the standard analysis of quantum
amplitudes fail, is presented.
\end{abstract}

\begin{keyword}
rectangular potential \sep Green's function \sep 
bound state \sep resonant scattering

\end{keyword}

\end{frontmatter}

\section{Introduction}
\label{sec:introduction}

One-dimensional systems are used frequently in quantum mechanics
to approximate real situations because they are relatively easy
to treat mathematically, allowing to get a deeper insight on the
physics involved.
They are useful in a number of applications in contemporary
physics and their simplicity has made them valuable as academic
and research tools.
For instance, it turns up in the description of resonant
tunneling  diode devices, disordered one-dimensional lattices,
realistic one-dimensional solid-state system such as quantum
wells, junctions and superlattices
\cite{PRL.1998.81.5796,PRB.1999.60.10569},
and time analysis of one-dimensional tunneling processes
\cite{RMP.1989.61.917}.
In particular, one-di\-men\-sion\-al rectangular potentials can be
used to model isolated transitions, observed in semiconductors,
from a bound state within a quantum well to a bound state at an
energy greater than the barrier height \cite{Nature.1992.358.565}.
Limiting cases of rectangular potentials can be used to describe
point interactions \cite{Book.1999.Flugge}, as well as
regularized singular interactions \cite{PR.1958.112.1137}.
Also, we should point out that the use of square-barrier and
square-well potentials as simple models for more realistic
physical problems has a long history in the theory of
heterostructures in solid-state physics \cite{Book.1991.Weisbuch}.

However, the propagator or its Fourier transform, the Green's
function, for rectangular potentials, although these are very
simple systems, are difficult to obtain and involve lengthy and
tedious calculations.
Indeed, the exact Green's function and propagators for step
potential and square barrier was obtained only in the late
1980s until the early 1990s
\cite{ZN.1985.40a.379,PRA.1992.46.1233,PRA.1993.47.2562,
PRA.1993.48.2567,PRA.1995.51.2654}.
Moreover, it is well known that semiclassical approach, i.e.,
the Van Vleck-Gutzwiller formula \cite{Book.1990.Gutzwiller}, or
the WKB approach \cite{Book.1998.Merzbacher}, give poor results
when applied to the class of rectangular potentials due to their
discontinuities.

In the present work, we show how to obtain the exact Green's
function
\cite{JPA.1998.31.2975,JPA.2001.34.5041,JPA.2003.36.227},
given as a general sum over paths, in a very simple way avoiding
complicated calculations, for arbitrary rectangular potentials
enlarging the knowledge of exact solutions for Green’s functions
and also generalizing and resuming results in the literature.
The procedure outlined here can be thought of as exact version
of the approximation in \cite{AJP.1983.51.897} and provides simple
and direct way to construct recurrence relations for quantum
amplitudes for one-dimensional scattering discussed in
\cite{PRA.1994.49.3310}.
Specifically, we focus on Green's function for single quantum
wells to avoid extra and unnecessary mathematical complications,
but the method is general and could be applied to multiple
quantum wells as well \cite{JPA.2003.36.227}.
From the Green's functions obtained, we  thus describe how
to extract the bound states eigenenergies and eigenfunctions.
Also, we discuss an application of the Green's function approach
to extract information of quasi-bound states in rectangular
potentials.
We should observe that the Green's function approach used in the
present work, has been successfully used in the calculation of
exact Green's functions for segmented potentials
\cite{JPA.1998.31.2975}, in calculation of asymptotic Green's
functions \cite{JPA.2001.34.5041}, in the determination and
discussion of bound states in multiple quantum wells
\cite{JPA.2003.36.227}.
Also, in the calculation of exact Green's function for quantum
graphs \cite{JPA.2003.36.545}, general point interactions
\cite{PRA.2002.66.062712}, and scattering quantum walks
\cite{PRA.2011.84.42343,PRA.2012.86.042309}.

This paper is organized as follows.
In Sec. \ref{sec:gfdp}, we give a briefly review on the
definition and properties of the Green's functions.
In Sec. \ref{sec:gf}, the Green's functions approach for the
rectangular potential is presented.
In Sec. \ref{sec:awp}, the construction of the exact Green's
functions for a general rectangular asymmetric well potential is
presented and how the sought eigenenergies and eigenfunctions are
extracted.
In Sec. \ref{sec:swp}, the case of the square well potential is
discussed and the definite parity eigenfunctions are determined.
In Sec. \ref{sec:iwp} the case of the infinite well potential is
discussed.
In Sec. \ref{sec:qbs} a Green's function approach to extract
information from system with quasi-bound states is presented.
Finally, Sec. \ref{sec:conclusion} contains our conclusions.

\section{Green's function: Definition and properties}
\label{sec:gfdp}

The Green's functions are an important tool in quantum mechanics
for calculating eigenenergies, eigenfunctions and density
of states \cite{Book.2006.Economou}.
It can be defined as solutions of the inhomogeneous differential
equation
\begin{equation}
  \label{eq:defgf}
  [E-\hat{H}(x_{f})]G(x_{f},x_{i};E)=
  \delta(x_{f}-x_{i}),
\end{equation}
subject to certain boundary conditions.
$\hat{H}(x)$ is the Hamiltonian operator
\begin{equation}
  \label{eq:Hamiltonian}
  \hat{H}(x)=-\frac{\hbar^{2}}{2m}\frac{d^{2}}{dx^{2}}+V(x).
\end{equation}

The Green's function $G(x_{f},x_{i};E)$ for a quantum system can be
interpreted as a probability amplitude for a particle that initially at
the point $x_{i}$ moves to the point $x_{f}$ for a fixed
energy \cite{Book.2005.Schulman}
\begin{equation}
G(x_{f},x_{i};E)=\langle x_{f}|\hat{G}(E)|x_{i} \rangle,
\end{equation}
where
$\hat{G}(E)\equiv \lim_{\eta \to 0^{+}} [E-\hat{H}(x)+i\eta]^{-1}$
is the resolvent operator \cite{Book.2006.Kleinert}.

Suppose we have a complete set of normalized energy eigenstates
$\psi_{n}(x)$, $n=0,1,...$ with
\begin{equation}
\hat{H}(x)\psi_{n}=E_{n} \psi_{n},
\end{equation}
so that the solution of Eq. \eqref{eq:defgf} can be written as
\begin{equation}
  \label{eq:segf}
  G(x_{f},x_{i};E)=
  \sum_{n}\frac{\psi_{n}(x_{f})\psi_{n}^{*}(x_ {i})}{E - E_n}.
\end{equation}
For a discrete spectrum, we thus identify the poles of
Green's function with bound states eigenenergies and the
residues at these points give a tensorial product of the bound
states eigenfunctions.
The  continuous part of the spectrum corresponds to a branch
cut of $G(x_{f},x_{i};E)$ \cite{JMP.1992.33.643,Book.2006.Kleinert}.
The following limit
\begin{equation}
\label{eq:limg}
\lim_{E \to E_n}(E-E_n)G(x_f,x_i;E)=\psi_n{(x_f)}\; \psi_n^{*}{(x_i)}
\end{equation}
is used to extract the discrete bound states from the
Green's function.

There are basically three methods for calculating the Green's
functions \cite{Book.2006.Economou}:
solving the differential equation in \eqref{eq:defgf};
summing up the spectral representation in \eqref{eq:segf};
or performing the Feynman path integral expansion for the
Green's function \cite{Book.2010.Feynman}.
In this paper, we present a Feynman-like procedure.
As we are going to see, the approach requires one to determine
the quantum amplitudes of each individual potential, but in
general, this is a much easier task than to calculate numerically
propagation along the whole $V(x)$.

\section{The Green's function approach for rectangular
  potentials}
\label{sec:gf}

\begin{figure}[t*]
    \centering
    \includegraphics*[width=0.85\columnwidth]{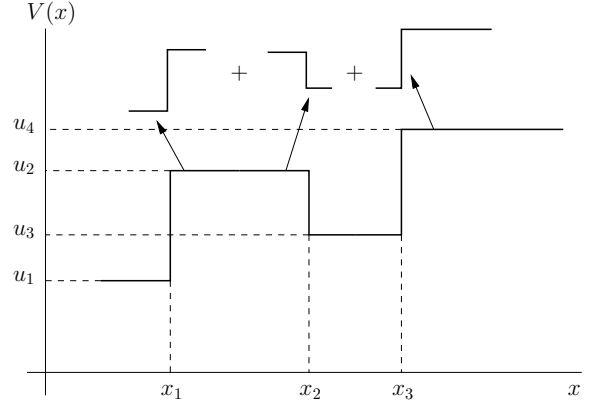}
    \caption{A rectangular potential with three points of
      discontinuities.
      The potential can be considered as a sum of the three
      potential steps.}
    \label{fig:fig1}
\end{figure}

According to Refs.
\cite{JPA.1998.31.2975,JPA.2001.34.5041,JPA.2003.36.227}
the Green's function $G(x_{f},x_{i};E)$ (in what follow, we will
denote it by $G_{f,i}$ for short) for a general 1D
potential is obtained by  writing the potential $V(x)$ as a sum of
$n$ individual potentials $u_{j}$, where each $u_{j}$ vanish
asymptotically.
In the present work, we focus our attention on rectangular
potentials.
As depicted in Fig. \ref{fig:fig1}, a rectangular
potential can be treated as the sum of $n$ step potentials.
The potential function for a rectangular potential can be
written as
\begin{equation}
  V(x)=\left\{
    \begin{array}{llll}
      u_1 & \text{for} & x < x_1,
      & \text{(Region $1$)},\\
      u_j & \text{for} & x_j\leq x < x_{j+1}=x_{j}+\ell_{j},
      & \text{(Region $j$)},\\
      u_n & \text{for} & x \geq x_{n},
      & \text{(Region $n$)},
    \end{array}
  \right.
  \label{eq:potpcp}
\end{equation}
where $u_{j}$, $\ell_j$ and $x_j$ are the height of the potential,
the length of the region $j$ and the localization of the $j$th
discontinuous point, respectively.
The wave number in each different region is
$k_j=\sqrt{2m(E-u_j)/\hbar}=i\kappa_j$.
In what follow, we do not distinguish the cases whether $k_{j}$
is a real or a purely imaginary number ($i\kappa_{j}$), except otherwise
mentioned.

The Green's function for this system, for a fixed energy $E$ and
the end points $x_{i}$ and $x_{f}$, is given by
\begin{equation}
  \label{eq:fgsg}
  G_{f,i}=\frac{m}{i\hbar^2\sqrt{k_{f}k_{i}}}
  \sum_{\rm{sp}}
  W_{\rm{sp}}
  \exp{[\frac{i}{\hbar} S_{\rm{sp}}(x_{f},x_{i};E)]},
\end{equation}
where $W_{\rm sp}$ is the amplitude (or weight) and
$S_{\rm sp}$ is the classical action.
The above sum is performed over all scattering paths (sp)
starting in $x_{i}$ and ending in $x_{f}$.
For each sp, the classical action is obtained from the
propagation over action of potential $u_{j}$,
$S_{\rm sp}(x_b,x_a;E)=\hbar \int_{x_a}^{x_b}k_{j} dx$.
A few comments concerning the Eq. \eqref{eq:fgsg} are necessary.
The expression in Eq. \eqref{eq:fgsg} is obtained from a recursive
procedure, i.e., $G_{f,i}$ for $n$ potentials are derived from
$G_{f,i}$ for $n-1$ potentials.
The $W_{\rm sp}$ are related to local quantum effects,
so they depend on quantum amplitudes ($R_{j}^{(\pm)}$ and
$T_{j}^{(\pm)}$) of individual $u_{j}$
(cf. \ref{sec:appendixa}).
In fact, for the present case of rectangular potentials,
$R_{j}^{(\pm)}$ and $T_{j}^{(\pm)}$ are the usual reflection
and transmission amplitudes of individual $u_{j}$
\cite{PRA.1994.49.3310} (up to a phase
\cite{JPA.1998.31.2975}).
(The superscript $(+)$ ($(-)$) and subscript $j$ denote the
physical quantities of a quantum particle incident
from left (right) at the point $x_{j}$, respectively.)
In the general case of smooth potentials, they are related, but
we need to calculate classical actions for each potential.
In this later case, the interested reader can consult the
Refs. \cite{JPA.2001.34.5041,JPA.2003.36.227} for derivations.

\section{Asymmetric well potential}
\label{sec:awp}

\begin{figure}[t*]
    \centering
    \includegraphics*[width=0.85\columnwidth]{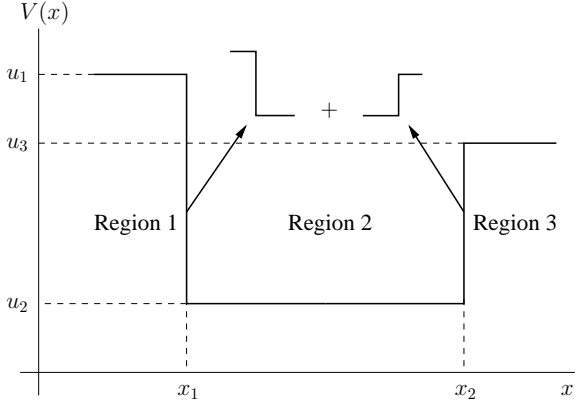}
    \caption{An asymmetric well potential written as a sum of a
      left and right potential steps.}
    \label{fig:fig2}
\end{figure}

In this section we will consider the case of a quantum particle
confined in the asymmetric well potential defined by
\begin{equation}
  \label{eq:aswp}
  V(x)=
  \left\{
  \begin{array}{llll}
    u_{1} & \text{for} & x < x_{1},
    & \text{(Region 1)},\\
    u_{2} & \text{for} & x_{1} \leq x<x_{2}=x_{1}+\ell_{2},
    &\text{(Region 2)},\\
    u_{3} & \text{for} & x \geq x_{2},
    &\text{(Region 3)},
  \end{array}
  \right.
\end{equation}
and depicted in Fig. \ref{fig:fig2}.
We seek the Green's function for the situation where
$E<u_{3}$.
The eigenvalues will be obtained from the poles of the Green's
function and the eigenfunctions from their respective residues
(cf. Sec. \ref{sec:gfdp}).
We can write nine different Green's functions for the problem,
but for our purposes, we analyze only three relevant
possibilities, namely:
(1) $x_{1}<x_{i}<x_{2}$ and $x_{f}<x_{1}$;
(2) $x_{1}<x_{i},x_{f}<x_{2}$; and
(3) $x_{1}<x_{i}<x_{2}$ and $x_{f}>x_{2}$.
The others six Green's functions can be obtained from symmetry
considerations of the three above cases.

First, we want to exemplify how to use the Eq. \eqref{eq:fgsg}
and how the geometric series appear.
So, let us consider the possibility (2).
To obtain $G_{f,i}^{(2)}$ in a closed form, we need to sum up
all the possibles scattering paths for a quantum particle
starting from $x_{i}$ and arriving at $x_f$, but it can always
be done because the sum in \eqref{eq:fgsg} forms a geometric
series. Without loss of generality, let us set $x_{f}>x_{i}$.

\begin{figure}[t*]
    \centering
    \includegraphics*[width=\columnwidth]{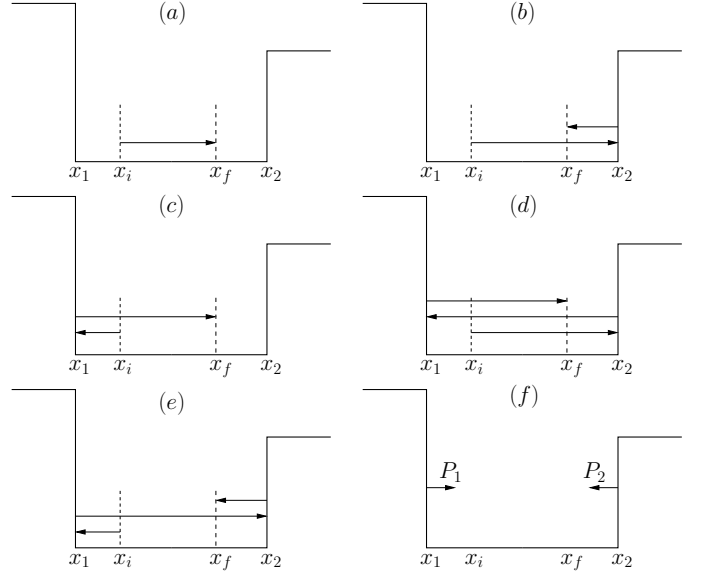}
    \caption{In (a)-(e) schematic examples of scattering paths
      and in (f) the families of paths $P_{1}$ and $P_{2}$.}
    \label{fig:fig3}
\end{figure}

Thus, in Fig. \ref{fig:fig3} is depicted five examples of
scattering paths. The first path in Fig. \ref{fig:fig3}(a) is
the direct propagation from $x_i$ to $x_f$, which contributes
with $e^{i k_{2} (x_f-x_i)}$ for $G_{f,i}^{(2)}$.
In Fig. \ref{fig:fig3}(b), a quantum particle leaves the point
$x_i$, goes to the point $x_{2}$, where it hit the potential and
is reflected, and so, goes to left to arrives at the point $x_f$.
This path contributes with
$e^{i k_{2}(x_{2}-x_{i})}R_{2}^{(+)}e^{-ik_{2}(x_{f}-x_{2})}$,
where $R_{2}^{(+)}$ is the reflection amplitude for a quantum
particle incident from the left of the potential $u_{3}$ at the
point $x_2$.
Following the same reasoning, the contributions for the other
examples in Fig. \ref{fig:fig3} are:
\begin{align*}
\text{(c) } {} &
e^{-ik_{2}(x_{1}-x_{i})}R_{1}^{(-)}e^{ik_{2}(x_{f}-x_{1})};
\nonumber \\
\text{(d) } {} &
e^{ik_{2}(x_{2}-x_{i})}R_{2}^{(+)}e^{ik_{2}\ell_{2}}
R_{1}^{(-)}e^{ik_{2}(x_f-x_{1})};
\nonumber \\
\text{(e) } {} &
e^{-ik_{2}(x_{1}-x_{i})}R_{1}^{(-)}e^{ik_{2}\ell_{2}}
R_{2}^{(+)}e^{-ik_{2}(x_{f}-x_{2})}.
\end{align*}

The Green's function is then given by the sum of all
contributions of $n$ multiple reflections of all possibles
scattering paths and can be written as ($\ell_{2}=x_{2}-x_{1}$)
\begin{align}
  G_{f,i}^{(2)}  =
  {} & \frac{m}{ik_{2}\hbar^2}
  \Big\{e^{ik_{2}(x_f-x_i)}
  \nonumber \\
  {} & +\sum_{n=0}^{\infty}
  \left(R_{1}^{(-)}R_{2}^{(+)} e^{2  i k_{2} \ell_{2}}\right)^{n}
  \Big[e^{-ik_{2}(x_{1}-x_{i})}R_{1}^{(-)}
   \nonumber \\
  {} & \times
  \left(e^{ik_{2}(x_{f}-x_{1})}+
  R_{2}^{(+)}e^{ik_{2}\ell_{2}}e^{-ik_{2}(x_{f}-x_{2})}\right)
  \nonumber \\
  {} & + e^{ik_{2}(x_{2}-x_{i})}R_{2}^{(+)}
  \nonumber \\
  {} & \times
  \left(e^{-ik_{2}(x_{f}-x_{2})}+
    R_{1}^{(-)}e^{ik_{2}\ell_{2}}e^{ik_{2}(x_{f}-x_{1})}\right)\Big]\Big\}.
\label{eq:gexefinal}
\end{align}
In general, the reflections (transmissions) amplitudes have the
property  $|R_{j}^{(\pm)}|^{2}\leq 1$ ($|T_{j}^{(\pm)}|^{2}\leq 1$),
in such way as the above sum always converge.
In fact, it is a geometric series.
So, after a straightforward algebra the final form for the
Green's function is
\begin{align}
  G_{f,i}^{(2)} =
  {} &\frac{m}{i\hbar^2 k_{2} f_{\rm aw}}
  \left(
    e^{i k_{2} (x_{f}-x_{1})}+
    R_{2}^{(+)}e^{i k_{2} \ell_{2}}
    e^{i k_{2} (x_{2}-x_{f})}
  \right)
  \nonumber \\
  {} & \times
  \left(
    e^{-ik_{2}(x_{i}-x_{1})}+
    R_{1}^{(-)}e^{i k_{2} \ell_{2}}
    e^{-ik_{2}(x_{2}-x_{i})}
  \right) ,
  \label{eq:gaswpcase2}
\end{align}
where $f_{\rm aw}=1-R_{1}^{(-)}R_{2}^{(+)}e^{2 i k_{2} \ell_{2}}$.

The method utilized above consists in sum up all scattering
paths, but this could be very tedious and cumbersome.
However, this can be done by a simple diagrammatic
classification in families of paths, which is a practical way to
identify and perform the geometric series mentioned above.
Consider again the confined particle in the asymmetric well
potential.
The particle starting from $x_{i}$ may
(a) go to the right arriving $x_{f}$;
or (b) go to the left, hit the potential $u_{1}$ at $x_{1}$
being reflected.
There is an infinite family of paths for the particle at
$x_{1}$, we call this family $P_{1}$;
and (c) the particle go to the right, hit the potential $u_{3}$
at $x_{2}$ being reflected.
Like before, there is infinite family of paths for the particle
at $x_{2}$, we call this family $P_{2}$.
These two infinite family of paths are schematically depicted in
Fig. \ref{fig:fig3}(f).
Thus, by using the above prescription, the Green's function can be
written as
\begin{align}
  G_{f,i}^{(2)}
  ={} &\frac{m}{i\hbar^2 k_{2}}
  \left(e^{i k_{2} (x_{f}-x_{i})}\right.
  \nonumber \\
  {} & +
  \left. e^{i k_{2} (x_{2}-x_{i})}R_{2}^{(+)}P_{2} +
  e^{-i k_{2} (x_{1}-x_{i})}R_{1}^{(-)}P_{1}\right).
\label{eq:gsw}
\end{align}
The family $P_{1}$ ($P_{2}$) posses two contributions:
(a) go to the right (left), arriving at $x_f$;
and (b) go to the right (left), hit the potential
$u_{3}$ ($u_{1}$) at $x_{2}$ ($x_{1}$) being reflected followed
by the family $P_{2}$ ($P_{1}$).
Thus,
\begin{subequations}
  \label{eq:Ps}
  \begin{align}
    P_{1}={} & e^{i k_{2} (x_{f}-x_{1})} +
    e^{i k_{2} \ell_{2}}R_{2}^{(+)}P_{2},
    \label{eq:Pls} \\
    P_{2}={} & e^{i k_{2} (x_{f}-x_{2})} +
    e^{i k_{2} \ell_{2}}R_{1}^{(-)}P_{1}.
    \label{eq:Prs}
  \end{align}
\end{subequations}
Solving for $P_{1}$ and $P_{2}$ one obtains
\begin{subequations}
  \label{eq:Psol}
  \begin{align}
    P_{1}={} & \frac{1}{f_{\rm aw}}\left(e^{ i k_{2} (x_{f}-x_{1})}+
      R_{2}^{(+)}e^{i k_{2} \ell_{2}} e^{i k_{2} (x_{2}-x_{f})}\right),
    \label{eq:Pl}\\
    P_{2}={} & \frac{1}{f_{\rm aw}}\left(e^{ i k_{2} (x_{2}-x_{f})}
      + R_{1}^{(-)}e^{i k_{2} \ell_{2}}e^{i k_{2}(x_{f}-x_{1})}\right).
    \label{eq:Pr}
  \end{align}
\end{subequations}
By substitution of Eq. \eqref{eq:Psol} into
Eq. \eqref{eq:gsw}, one obtains the Green's function in
Eq. \eqref{eq:gaswpcase2}.
The other two cases can be obtained in a similar way.
For instance, for case (1) we have
\begin{align}
  \label{eq:gaswpcase1}
  G_{f,i}^{(1)}=
  {} &
  \frac{m}{i\hbar^2 \sqrt{k_{1} k_{2}} f_{\rm aw}}
  \left(
    T_{1}^{(-)} e^{-i k_{1} (x_f-x_{1})}
  \right)
  \nonumber \\
  {} &
  \times
  \left(
    e^{-i k_{2} (x_i-x_{1})} + R_{1}^{(-)} e^{i k_{2} \ell_{2}}
    e^{-i k_{2} (x_{2}-x_i)}
  \right),
\end{align}
and for case (3), we obtain
\begin{align}
  \label{eq:aswpcase3}
  G_{f,i}^{(3)}=
  {} &
  \frac{m}{i\hbar^2 \sqrt{k_{3} k_{2}} f_{\rm aw}}
  \left(
    T_{2}^{(+)} e^{i k_{3}(x_f-x_{2})}
  \right)
  \nonumber \\
  {} & \times
  \left(
    e^{-i k_{2} (x_i-x_{1})} + R_{1}^{(-)}
    e^{ i k_{2} \ell_{2}}e^{-i k_{2} (x_{2}-x_i)}
  \right).
\end{align}

\subsection{Calculation of the bound states}

The bound states are calculated from the residues of the Green's
functions in Eqs. \eqref{eq:gaswpcase2}, \eqref{eq:gaswpcase1}
and \eqref{eq:aswpcase3}.
Its poles $E_{n}=\hbar^{2} k_{n}^{2}/2m$ are all contained in
the term $1/f_{\rm aw}$.
They are calculated from $f_{\rm aw}=0$, which leads to the
following transcendental equation
\begin{equation}
  \label{eq:aswpoles}
  \left(\frac{k_{2}^{(n)}-k_{1}^{(n)}}{k_{1}^{(n)}+k_{2}^{(n)}}\right)
  \left(\frac{k_{2}^{(n)}-k_{3}^{(n)}}{k_{2}^{(n)}+k_{3}^{(n)}}\right)
  e^{2 i k_{2}^{(n)} \ell_{2}}=1,
\end{equation}
where $k_{j}^{(n)}=\sqrt{2m(E_{n}-u_{j})}/\hbar$.
This result agree with the one found by the solution of the
Schr\"{o}dinger equation \cite{Book.1981.Landau}.
Using the formula
\begin{equation}
  \label{eq:limpolo}
  \lim_{E \to E_{n}}\frac{E-E_{n}}{f_{\rm aw}}=
  \frac{\hbar^{2}}{2m}
  \lim_{k \to k_{n}}\frac{k^{2}-k_{n}^{2}}{f_{\rm aw}}=
  \frac{\hbar^{2}}{m} \frac{k_{n}}{f_{\rm aw}^{'(n)}},
\end{equation}
where $f_{\rm aw}^{'(n)}=(d f_{\rm aw}/dk)|_{k=k_{n}}$, we obtain for
the residues of the Green's function in Eqs. \eqref{eq:gaswpcase2},
\eqref{eq:gaswpcase1} and Eq. \eqref{eq:aswpcase3}
\begin{align*}
  \makebox[0.5cm][l]{$\displaystyle
  \psi_{n}^{(2)}(x_f) [\psi_{n}^{(2)}(x_i)]^{*}=
  \frac{\hbar^{2}}{2m}
  \lim_{k \to k_{n}}(k^{2}-k_{n}^{2})G_{f,i}^{(2)}$} &
\nonumber \\
  = {} &\frac{1}{i f_{\rm aw}^{'(n)}}
  \left[
    e^{i k_{2}^{(n)} (x_{f}-x_{1})}+
    \frac{k_{2}^{(n)}-k_{3}^{(n)}}{k_{2}^{(n)}+k_{3}^{(n)}}\;
    e^{i k_{2}^{(n)} (\ell_{2} +x_{2}-x_{f})}
  \right]\nonumber \\
 {} & \times
  \left[
    e^{-ik_{2}^{(n)}(x_{i}-x_{1})}+
    \frac{k_{2}^{(n)}-k_{1}^{(n)}}{k_{1}^{(n)}+k_{2}^{(n)}}\;
    e^{i k_{2}^{(n)}(\ell_{2}-x_{2}+x_{i})}
  \right],
\end{align*}
\begin{align*}
  \makebox[0.6cm][l]{$\displaystyle
  \psi_{n}^{(1)}(x_f) [\psi_{n}^{(2)}(x_i)]^{*}=
  \frac{\hbar^{2}}{2m}\lim_{k \to
    k_{n}}(k^{2}-k_{n}^{2})G_{f,i}^{(2)}$} & \nonumber \\
  = {} & \frac{1}{i f_{\rm aw}^{'(n)}}
  \left[
      \sqrt{\frac{k_{1}^{(n)}}{k_{2}^{(n)}}}
      \frac{2k_{2}^{(n)}}{k_{1}^{(n)}+k_{2}^{(n)}}
    e^{-i k_{1}^{(n)} (x_f-x_{1})}
  \right] \nonumber \\
  {} & \times
  \left[
    e^{-i k_{2}^{(n)} (x_i-x_{1})} +
    \frac{k_{2}^{(n)}-k_{1}^{(n)}}{k_{1}^{(n)}+k_{2}^{(n)}}\;
    e^{i k_{2}^{(n)} (\ell_{2}-x_{2}+x_i)}
  \right],
\end{align*}
\begin{align*}
  \makebox[0.7cm][l]{$\displaystyle\psi_{n}^{(3)}(x_f)
    [\psi_{n}^{(2)}(x_i)]^{*}=
  \frac{\hbar^{2}}{2m}\lim_{k \to
    k_{n}}(k^{2}-k_{n}^{2})G_{f,i}^{(3)}$} & \nonumber \\
  = {} & \frac{1}{i f_{\rm aw}^{'(n)}}
  \left[
      \sqrt{\frac{k_{3}^{(n)}}{k_{2}^{(n)}}}
      \frac{2 k_{2}^{(n)}}{k_{2}^{(n)}+k_{3}^{(n)}}\;
    e^{i k_{3}^{(n)}(x_f-x_{2})}
  \right]\nonumber \\
  {} & \times
  \left[
    e^{-i k_{2}^{(n)} (x_i-x_{1})} +
    \frac{k_{2}^{(n)}-k_{1}^{(n)}}{k_{1}^{(n)}+k_{2}^{(n)}}\;
    e^{i k_{2}^{(n)} (\ell_{2}-x_{2}+x_i)}
  \right].
\end{align*}
The correct normalized eigenfunctions corresponding to the bound
states of the system are thus given by:\\
Region 1: $x<x_{1}$
\begin{equation}
  \psi_{n}^{(1)}(x)=
    \sqrt{\frac{1}{i f_{\rm aw}^{'(n)}}
      \frac{k_{1}^{(n)}}{k_{2}^{(n)}}}
        \frac{2k_{2}^{(n)}}{k_{1}^{(n)}+k_{2}^{(n)}}\;
      e^{-i k_{1}^{(n)} (x-x_{1})},
\end{equation}
Region 2: $x_{1} < x < x_{2}$
\begin{align}
  \psi_{n}^{(2)}(x)=
  {} &
  \sqrt{\frac{1}{i f_{\rm aw}^{'(n)}}}
  \Big(
    e^{i k_{2}^{(n)} (x-x_{1})}
    \nonumber \\
   {} & +
     +\frac{k_{2}^{(n)}-k_{3}^{(n)}}{k_{2}^{(n)}+k_{3}^{(n)}}\;
  e^{i k_{2}^{(n)} (\ell_{2} +x_{2}-x)}
  \Big),
\end{align}
Region 3: $x>x_{2}$
\begin{equation}
  \psi_{n}^{(3)}(x)=
  \sqrt{\frac{1}{i f_{\rm aw}^{'(n)}}
    \frac{k_{3}^{(n)}}{k_{2}^{(n)}}}
  \frac{2 k_{2}^{(n)}}{k_{2}^{(n)}+k_{3}^{(n)}}\;
  e^{i k_{3}^{(n)}(x-x_{2})}.
\end{equation}
It should be observed that $k_{2}^{(n)}$
($k_{1}^{(n)}$ and $k_{3}^{(n)}$) is (are) real (imaginary) number(s),
and consequently $\psi_{n}^{(2)}(x)$ ($\psi_{n}^{(1)}(x)$ and
$\psi_{n}^{(3)}(x)$) is (are) an oscillatory complex (decreasing
real) exponential function(s) of $x$.
Also, the Green's function used here yields the correct
normalization constant which often involves a difficult integral
in the other methods.
In what follow, we apply the results obtained in this section in
other well-known rectangular potentials.

\section{Square well potential}
\label{sec:swp}

\begin{figure}[t*]
    \centering
    \includegraphics*[width=0.85\columnwidth]{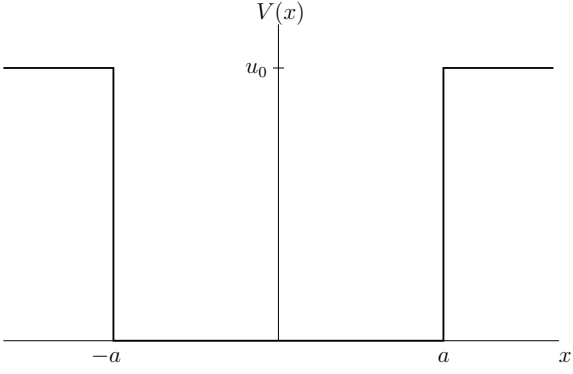}
    \caption{A square well potential.}
    \label{fig:fig4}
\end{figure}

The Green's function for a square well potential
(depicted in Fig. \ref{fig:fig4})
can be obtained from the results of the previous section by
setting $u_1=u_3=u_0$ and $u_{2}=0$, thus $k_1=k_3=k_0$ and
$k_{2}=k=\sqrt{2mE}/\hbar$, also $x_1=-a$ and $x_{2}=a$.
From Eqs. \eqref{eq:gaswpcase2}, \eqref{eq:gaswpcase1} and
\eqref{eq:aswpcase3} we thus find
\begin{align}
  \label{eq:swpcase1}
  G_{f,i}^{(1)}=
  {} &
  \frac{m}{i\hbar^2 \sqrt{k_{0} k} f_{\rm sw}}
  \left(
    T_{(-a)}^{(-)} e^{-i k_{0} (x_f+a)}
  \right)
  \nonumber \\
  {} &
  \times
  \left(
    e^{-i k (x_i+a)} + R_{(-a)}^{(-)} e^{2 i k a}
    e^{-i k (a-x_i)}
  \right),
\end{align}
\begin{align}
  \label{eq:sqwcase2}
  G_{f,i}^{(2)}=
  {} &
  \frac{m}{i\hbar^2 k f_{\rm sw}}
  \left(
    e^{i k (x_f+a)} + R_{(a)}^{(+)} e^{2 i k a} e^{i k (a-x_f)}
  \right)
  \nonumber \\
  {} & \times
  \left(
    e^{-i k (x_i+a)} + R_{(-a)}^{(-)} e^{2 i k a}e^{-i k (a-x_i)}
  \right),
\end{align}
\begin{align}
  \label{eq:swpcase3}
  G_{f,i}^{(3)}=
  {} & \frac{m}{i\hbar^2 \sqrt{k_{0} k} f_{\rm sw}}
  \left(
    T_{(a)}^{(+)} e^{i k_{0}(x_f-a)}
  \right)
  \nonumber \\
  {} & \times
  \left(
    e^{-i k (x_i+a)} + R_{(-a)}^{(-)} e^{ 2 i k a}e^{-i k (a-x_i)}
  \right),
\end{align}
where $f_{\rm sw}=1-R_{(-a)}^{(-)}R_{(a)}^{(+)} e^{4ika}$.

\subsection{Calculation of the bound states}

The eigenenergies are calculated from the condition
$f_{\rm sw}=0$.
By considering the symmetry of the potential, we can infer that
the reflections amplitudes are equal (up to a phase), i.e.,
$R_{(-a)}^{(-)}=R_{(a)}^{(+)}=(k-k_{0})/(k+k_{0})$,
leading to the following transcendental equation
\begin{equation}
  \label{eq:poloswp}
\left(\frac{k_{n}-k_{0}^{(n)}}{k_{n}+k_{0}^{(n)}}\right)^{2}
e^{4 i k_{n} a}=1.
\end{equation}
Using Eq. \eqref{eq:limpolo} the residues are determined and
leads to the following bound state eigenfunctions:\\
Region 1: $x<-a$
\begin{equation}
  \psi_{n}^{(1)}(x)=
    \sqrt{\frac{1}{i f_{\rm sw}^{'(n)}}
      \frac{k_{0}^{(n)}}{k_{n}}}
        \frac{2k_{n}}{k_{0}^{(n)}+k_{n}}\;
      e^{-i k_{0}^{(n)} (x+a)},
\end{equation}
Region 2: $-a < x < a$
\begin{align}
  \psi_{n}^{(2)}(x)=
  {} &
  \sqrt{\frac{1}{i f_{\rm sw}^{'(n)}}}
  \Big(
    e^{i k_{n} (x+a)}
    \nonumber \\
   {} &
     +\frac{k_{n}-k_{0}^{(n)}}{k_{n}+k_{0}^{(n)}}\;
     e^{2i k_{n} a}e^{ik_{n}(a-x)}
  \Big),
\end{align}
Region 3: $x>a$
\begin{equation}
  \psi_{n}^{(3)}(x)=
  \sqrt{\frac{1}{i f_{\rm sw}^{'(n)}}
    \frac{k_{0}^{(n)}}{k_{n}}}
  \frac{2 k_{n}}{k_{n}+k_{0}^{(n)}}\;
  e^{i k_{0}^{(n)}(x-a)}.
\end{equation}

From Eq. \eqref{eq:poloswp}, we can see that there are two
equations describing the eigenenergies.
These eigenenergies are determined by solving
\begin{align*}
(k_{n}-k_{0}^{(n)})e^{2 i k_{n} a}& = +(k_{n}+k_{0}^{(n)}), &
\text{(even parity)}, \\
(k_{n}-k_{0}^{(n)})e^{2 i k_{n} a}& = -(k_{n}+k_{0}^{(n)}), &
\text{(odd parity)},
\end{align*}
and the respective eigenfunctions are
\begin{equation}
  \psi_{n}^{(2)}(x)=2\sqrt{\frac{1}{i f_{\rm sw}^{'(n)}}}
  \begin{cases}
    e^{i k_{n} a} \cos{[k_{n} x]},   & \text{(even parity)}, \\
    i e^{i k_{n} a} \sin{[k_{n} x]}, & \text{(odd parity)},
  \end{cases}
\end{equation}
which are the well-known even and odd solutions for the square
well potential \cite{Book.2011.Sakurai}.

\section{Infinite well potential}
\label{sec:iwp}

In this section, we consider the infinite well potential.
Similarly as in previous sections, the Green's function for the
infinite well potential can be obtained by letting
$u_{1}=u_{3}=\infty$ and $x_1=0$ and $x_2=L$, so that the
potential function is given by
\begin{equation}
  V(x)=
  \begin{cases}
    0  & \text{for } 0 < x < L, \\
    \infty & \text{otherwise}.
  \end{cases}
\end{equation}
As the potentials are infinite in $x_1$ and $x_2$, the
reflection amplitude assume the value
$R_{(0)}^{(-)}=R_{(L)}^{(+)}=-1$
and a vanishing transmission amplitude.
Thus, the Green's function in Regions 1 and 3 vanishes, so
that $\psi_{n}^{(1)}(x)= \psi_{n}^{(3)}(x)=0$.
The Green's function in Region 2 take the form
\begin{align}
  \label{eq:giswp}
  G_{f,i}^{(2)}=
  {} &
  \frac{m}{i\hbar^2 k f_{\rm iw}}
  \left(
    e^{i k x_f}-e^{i k L}e^{i k (L-x_f)}
  \right)
  \nonumber \\
  {} & \times
  \left(
    e^{-i k x_i}-e^{i k L}e^{-i k (L-x_i)}
  \right),
\end{align}
with $f_{\rm iw}=1-e^{2 i k L}$.
After a straightforward algebra, the Eq. \eqref{eq:giswp} can be
written as
\begin{equation}
  G_{f,i}^{(2)}=
  \frac{2m}{\hbar^2 k \sin{[k L]}}
  \sin{[k(x_{f}-L)]\sin{[k x_{i}]}}.
\end{equation}
This result is indeed the exact one obtained through the
spectral expansion of the Green's function in
Eq. \eqref{eq:segf} (cf. \ref{sec:appendixb}).

\subsection{Calculation of the bound states}

The poles are all contained in $\sin{[kL]}$, and eigenenergies of
the  bound states are thus obtained from
$\sin{[k_{n}L]}=0$, i.e.,
\begin{equation}
k_{n}=\frac{n\pi}{L}, \quad n=1,2,3,\ldots.
\end{equation}
From the residues of the Green's function, the correctly
normalized eigenfunctions corresponding for the bound states are
thus given by
\begin{equation}
\psi_{n}^{(2)}(x)= \sqrt{\frac{2}{L}}
  \sin{\left(\frac{n \pi x}{L}\right)}.
\end{equation}
Indeed, these are the exact results for the bound states for the
infinite well potential.

\section{Application on quasi-bound states}
\label{sec:qbs}

In this section, we apply the Green's function approach of
previous sections to extract information of systems with
quasi-bound states.
A quasi-bound state occurs when a particle move inside a system
for a considerable period of time, leaving it when a fairly long
time interval $\tau$ has elapsed \cite{Book.1981.Landau}, where
$\tau$ is called lifetime of the quasi-bound state.
The concept of quasi-bound states is a fundamental one, and has
been applied in all areas of physics.
They have been used to calculate tunneling ionization rates
\cite{PRL.1998.81.2663}, to understand the phenomenon of
diffraction in time \cite{PRA.2011.83.043608}, to describe the
decay of cold atoms in quasi-one-dimensional traps
\cite{EPL.2006.74.965}, and are directly relevant to recent
condensed-matter experiments \cite{PRL.2005.95.066801}.

\begin{figure}[t*]
    \centering
    \includegraphics*[width=0.9\columnwidth]{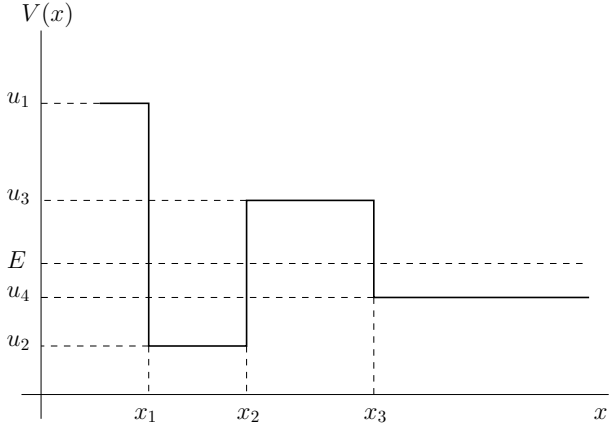}
    \caption{A rectangular potential able to support quasi-bound
      states as discussed in the text.}
    \label{fig:fig5}
\end{figure}

Let us consider the potential depicted in Fig. \ref{fig:fig5}.
Suppose the potentials $u_1$ and $u_3$ were infinitely high.
It would then be possible for particles to be trapped inside the
region for $x_1<x<x_2$, i.e., the system  would have genuine
bound states, with well defined energy $E>0$.
They are genuine bound states in the sense that they are
eigenstates of the Hamiltonian with an infinite lifetime.
From the Heisenberg uncertainty principle,
$\Delta E \Delta t \approx \hbar$, if the energy has null
uncertainty its state's lifetime is infinite.

\begin{figure}[t*]
    \centering
    \includegraphics*[width=\columnwidth]{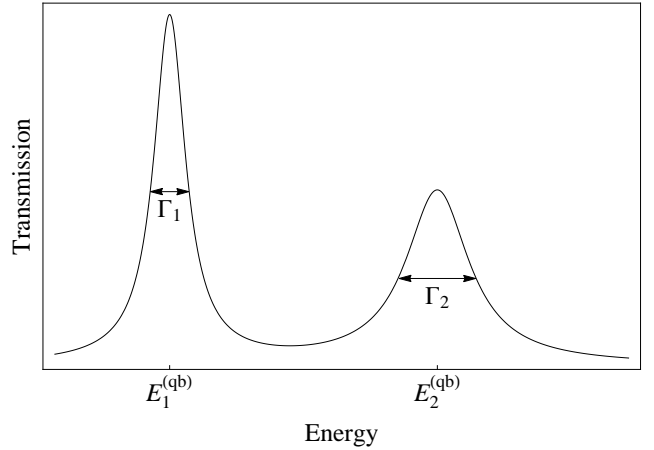}
    \caption{Typical behavior for the transmission coefficient
      for a potential which has two quasi-bound states with
      energies $E_{1}^{(\rm qb)}$ and $E_{2}^{(\rm qb)}$ and widths
      $\Gamma_1$ and $\Gamma_2$, respectively.}
    \label{fig:fig6}
\end{figure}

In the situation of a finite barrier as in Fig. \ref{fig:fig5}
(this is rough rectangular approximation for the effective
potential in a central force problem  $V(r)$ plus the
centrifugal barrier $(\hbar^{2}/2m) [l(l+1)/r^{2}]$
\cite{Book.2011.Sakurai}),
the particle can be trapped, but it cannot be trapped forever,
even if $E<u_{3}$, as a consequence of the tunnel effect.
The energy spectrum of these particle will be quasi-discrete, and
it consists of a series of broadened levels, whose width in
represented by $\Gamma=\hbar/\tau$ \cite{Book.1998.Merzbacher},
and the energy values are called quasi-energies.
In the scattering of particles by such potential, the situation
becomes very interesting when the incident energy is close to
the quasi-energy
\begin{equation}
E^{(\rm inc)}\approx E^{(\rm qb)}.
\end{equation}
In this energy interval, the module square of the transmission
amplitude exhibits pronounced peaks, and this is called resonant
scattering \cite{Book.2011.Sakurai}.
In Fig. \ref{fig:fig6} it is depicted a typical transmission probability
as a function of incident energy for a scattering of a potential which
supports quasi-bound states.

Now, let us consider the case of a finite square barrier at $x_2$
with an infinite barrier at $x_{1}$.
In this situation, the system can also has quasi-bound states,
due the tunneling through the right square barrier.
The scattering eigenfunction for a particle incident from the
right is given by
\begin{equation}
  \psi (x) \approx
  \frac{1}{\sqrt{2\pi}}(e^{-ik_{4} x}+R_{\rm pot}^{(-)} e^{i k_{4}x}),
\end{equation}
where $R_{\rm pot}^{(-)}$ is the reflection amplitude of whole
potential.
By analogy with the previous case, we would try to extract the
information from quasi-bound states from the reflection
coefficient $R_{\rm pot}^{(-)}$.
Unfortunately, due to the potential to be infinite at left, the
reflection coefficient has the value $|R_{\rm pot}^{(-)}|^{2}=1$ for
all range of energies.
Thus, we cannot extract information of quasi-bound states for
this kind of potential by the above method.
So, we propose a Green's function approach to extract
information of quasi-bound states for such kind of potential,
as we explain below.
Following the same steps described in the Section \ref{sec:awp},
the Green's function for $x_{i}>x_{3}$ and $x_{1}<x_{f}<x_{2}$
is readily obtained and is written as
\begin{align}
  \label{eq:gqbs}
  G_{f,i}=
   {} &
  \frac{m}{i\hbar^2 \sqrt{k_{2} k_{4}}}
  \frac{T_{b}^{(-)}}{f_{\rm qb}} e^{i k_{4} (x_{i}-x_{3})}
  \nonumber \\
  {} & \times
  \left(
    e^{ik_{2}(x_{2}-x_{f})}+R_{1}^{(-)}
    e^{ik_{2}(x_{f}+x_{2}-2x_{1})}
  \right),
\end{align}
with ${f}_{\rm  qb}=1-R_1^{(-)}R_{b}^{(+)}e^{ik_{2}\ell_{2}}$.
$R_{b}^{(+)}$ and $T_{b}^{(-)}$ are the reflection and transmission
amplitudes for the potential barrier given, respectively, by
\begin{equation}
  \label{eq:refb}
  R_{b}^{(+)}=
  R_{2}^{(+)}+
  \frac
  {T_{2}^{(-)}T_{2}^{(+)}R_{3}^{(+)}e^{2ik_{3}\ell_{3}}}
  {1-R_{2}^{(-)}R_{3}^{(+)}e^{2ik_{3}\ell_{3}}},
\end{equation}
\begin{equation}
  \label{eq:transb}
  T_{b}^{(-)}=
  \frac
  {T_{2}^{(-)}T_{3}^{(-)}e^{ik_{3}\ell_{3}}}
  {1-R_{2}^{(-)}R_{3}^{(+)}e^{2ik_{3}\ell_{3}}}.
\end{equation}

\begin{figure}[t*]
    \centering
    \includegraphics*[width=\columnwidth]{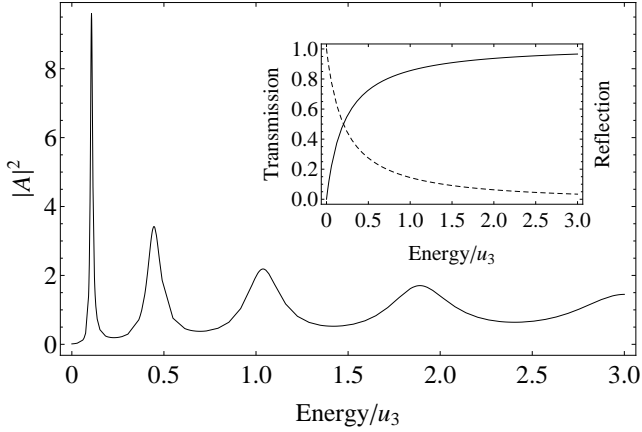}
    \caption{Behavior of $|A|^{2}=|T_{b}^{(-)}/\bar{f}_{qb}|^{2}$ as
      function of the energy displaying the presence of
      quasi-bound states.
      The parameters are $u_{1}=\infty$, $u_2=u_{4}=0$,
      $u_{3}=0.23$ eV, $\ell_{3}=80$ \AA {} and $\ell_{2}=10.05
      \ell_{3}$.
      In the inset is shown the transmission (solid line) and
      reflection (dashed line) coefficients for the potential
      barrier.
    }
    \label{fig:fig7}
\end{figure}

For an infinite barrier at $x_{1}$, we have $R_{1}^{(-)}=-1$,
and the Eq. \eqref{eq:gqbs} simplifies to
\begin{equation}
  \label{eq:gqbss}
  G_{f,i}=
  \frac{2m}{i^{2}\hbar^2 \sqrt{k_{2} k_{4}}}
  \frac{T_{b}^{(-)}}{\bar{f}_{\rm qb}} e^{i k_{4}(x_{i}-x_{3})}
  \sin[k_{2}x_{f}],
\end{equation}
where $\bar{f}_{\rm  qb}=1+R_{b}^{(+)}e^{ik_{2}\ell_{2}}$.
From the interpretation of the Green's function as a probability
amplitude (cf. Section \ref{sec:gfdp}), we can thus interpret
the term
\begin{equation}
 A=\frac{T_{b}^{(-)}}{\bar{f}_{\rm qb}},
\end{equation}
in Eq. \eqref{eq:gqbss} as such amplitude.
If the potential has quasi-bound states, an incident wave with
energy close to the quasi-energy, will have a high probability
of tunneling, entering in the trapping region.
Consequently, a graph of $|A|^{2}$ as a function of the energy,
will have peaks at each energy value close to
$E^{(\rm qb)}$.
So, we can extract information of quasi-energies and its
respective widths from $A$.
It is worthwhile to observe that the amplitude $A$ is not
normalized, but this is not a problem, because we are only
interested in the position of quasi-energies and the width of
quasi-states.

For a numerical example, in the Fig. \ref{fig:fig7} is
shown  a graph of $|A|^{2}$ as function of energy for the
potential in Fig. \ref{fig:fig5}.
For the parameters we choose typical values for heterostructures
in GaAs \cite{JPCM.1989.1.9027}.
The particle's mass is $m=0.07 m_{e}$, where $m_e$ is the
electron mass, $u_{1}=\infty$, $u_2=u_{4}=0$, $u_{3}=0.23$ eV,
$\ell_{3}=80$ \AA {} and $\ell_{2}=10.05 \ell_{3}$.
As we can see, it is evident in the graph the existence of
quasi-bound states.

The approach presented here for rectangular potentials, can be
generalized for smooth potentials, but in this case it is
necessary the calculation of the classical action for the
quantum particle under the action of the potential.
Specifically, we write the amplitude as
\begin{equation}
A=\frac{T_{b}}{f},
\end{equation}
where $T_{b}$ is the generalized transmission amplitude for the
smooth barrier between $x_{2}$ and $x_{3}$,
\begin{equation}
f=1-R R_{\infty} \exp{[\frac{i}{\hbar}S(x_{2},x_{3};E)]},
\end{equation}
where $S(x_{2},x_{3},k)$ is the classical action, $R$ is the
generalized reflection amplitude for the barrier, and
$R_{\infty}$ is the generalized reflection amplitude for the
infinity  barrier at $x_{1}$.
Since $|R_{\infty}|^{2}=1$, we can write
$R_{\infty}=\exp{[-i\phi(E)]}$, in such way that
\begin{equation}
f=1-R \exp{[\frac{i}{\hbar}S(x_{2},x_{3};E)-i\phi(E)]}.
\end{equation}
All those generalized amplitudes above are obtained by the procedure
outlined in \cite{JPA.2001.34.5041,JPA.2003.36.227}.

\section{Conclusion}
\label{sec:conclusion}

In this work, the exact Green's functions for rectangular single
wells are obtained in a rather general way and by a simple
method.
Our results are the exact ones and, although of the simplicity
of the systems considered, Green's functions for such system are
not so easy to obtain by standard procedures (for example,
solving the inhomogeneous differential equation in
Eq. \eqref{eq:defgf}).
The procedures allows one to discuss complete arbitrary
rectangular single wells and barriers, generalizing and resuming
results in the literature.
For instance, by withdrawing of the potential step at $x_{1}$ by
setting $R_{1}^{(-)}=0$, from Eq. \eqref{eq:gaswpcase2} the Green's
function for square barrier of Ref.
\cite{PRA.1993.48.2567,PRA.1995.51.2654}
is obtained.
The method can be applied for general potentials, including those
multidimensional with radial symmetry, but in this case is necessary
the calculation of the classical action for the particle under the
action of the potential \cite{JPA.2001.34.5041,JPA.2003.36.227}.

From the poles and residues of the Green's function the bound
state eigenenergies and eigenfunctions were obtained with the
correct normalization constant.
The determination of the later often involves a difficult
integral in the other methods.

Finally, we also have discussed an application of the Green's
function approach to extract information from quasi-bound states
in systems which standard analysis of the quantum amplitudes
are not possible.
The method could be generalized for smooth potentials and applied to the
well-known alpha decaying and determination of the dwell times and will
be subject of a future work.

\section*{Acknowledgments}

The author would like to thank E. O. Silva, C. F. Woellner and
J. A. O. Freire, for critical reading the manuscript and helpful
discussions.
This work was partially supported by the Fun\-da\-\c{c}\~{a}o
Arauc\'{a}ria (Brazil) under grant number No. 205/2013.

\appendix
\section{The quantum amplitudes for a step potential}
\label{sec:appendixa}

The step potential is used as building block of our
construction.
So, in this appendix we just outline the derivation of quantum
amplitudes for the step potential.
The potential function for the potential step is given by
\begin{equation}
 V(x) =
  \begin{cases}
    u_{j}    & \text{for } x < 0, \\
    u_{j+1}   & \text{for } x > 0.
  \end{cases}
\end{equation}
The reflection and transmission amplitudes are obtained from
solution of Schr\"{o}dinger equation.
The scattering solutions for the step potential with the
incident beam coming from $x=-\infty$ are
\begin{equation}
  \psi(x)=
    \begin{cases}
      e^{i k_{j} x}+ R_{j}^{(+)} e^{- i k_{j} x} & \text{for } x < 0,\\
      T_{j}^{(+)} e^{- ik_{j+1}x} & \text{for } x > 0,
    \end{cases}
\end{equation}
with $k_{j}=\sqrt{2 m (E-u_{j})}/\hbar$.
From the matching conditions at the origin, i.e.,
$\psi(0^{-})=\psi(0^{+})$ and $\psi'(0^{-})=\psi'(0^{+})$,
we find the sought reflection and transmission amplitudes
\begin{equation}
  \label{eq:Rsqw1}
  R_{j}^{(+)}=\frac{k_{j}- k_{j+1}}{k_{j}+k_{j+1}}, \qquad
  T_{j}^{(+)}=\sqrt{\frac{k_{j+1}}{k_{j}}}
  \frac{2 k_{j}}{k_{j}+k_{j+1}}.
\end{equation}
In the same way, for the case with the incident beam coming from
$x=\infty$, we have
\begin{equation}
  \label{eq:Rsqw2}
  R_{j}^{(-)}=-R_{j}^{(+)}, \qquad
  T_{j}^{(-)}=\sqrt{\frac{k_{j}}{k_{j+1}}}\frac{2 k_{j+1}}{k_{j}+k_{j+1}}.
\end{equation}
For the case $E>u_{j+1}$ ($E<u_{j+1}$), $k_{j+1}$ is a real
(imaginary) number and $T_{j}^{(\pm)}$ represents the transmission
(penetration) amplitude.

\section{Green's function for the infinite well potential
from spectral expansion}
\label{sec:appendixb}

In this appendix, we will calculate the Green's function for the
infinite well potential from the spectral expansion in
Eq. \eqref{eq:segf}.
The eigenfunctions and energies for a particle in the infinite
well potential are given by
\begin{align}
  \label{eq:wfisw}
  \psi_{n}= {} & \sqrt{\frac{2}{L}}
  \sin{\left(\frac{n \pi x}{L}\right)},
  \quad n=1,2,3,\ldots,
  \\
  E_{n}= {} &\frac{n^{2}\pi^{2}\hbar^{2}}{2 m L^{2}}.
\end{align}
Substituting the eigenfunctions in \eqref{eq:wfisw} into the
spectral expansion of the Green's function, Eq. \eqref{eq:segf},
we arrive at
\begin{equation}
  G_{f,i} =-\frac{4mL}{\hbar^2\pi^2}
  \sum_{n=1}^{\infty}
  \frac
  {\sin{\left({n X_{f}}\right)}
    \sin{\left({n X_{i}}\right)}}
  {n^2+\alpha^2},
\end{equation}
where $ X_{j}={\pi x_{j}}/{L}$, $\alpha^2=-{L^2k^2}/{\pi^2}$ and 
where $k=\sqrt{2mE}/\hbar$.
Using a trigonometric identity for the product of sines,
we have
\begin{align}
  G_{f,i}=
  {} & -\frac{2mL}{\hbar^2\pi^2}
  \bigg\{
  \sum_{n=1}^{\infty}
  \frac
  {\cos{[n(X_{f}-X_{i})]}}
  {n^2+\alpha^2}
  \nonumber \\
  {} &
  +\sum_{n=1}^{\infty}
  \frac
  {\cos{[n(X_{f}+X_{i})]}}
  {n^2+\alpha^2}
  \bigg\}.
\end{align}
The infinite sum above can be evaluated by using the identity
1.445-2 of Ref. \cite{Book.2007.Gradshteyn}, and after a
straightforward algebra we achieve at
\begin{equation}
  G_{f,i}=
  -\frac{2mL}{\hbar^2\pi \alpha \sinh{[\alpha \pi]}}
  \sinh{[\alpha(\pi-X_{f})]}\sinh{[\alpha X_{i}]}.
\end{equation}
Now, by substitution of $X_{j}$ and $\alpha$ and using
$\sinh{[i\theta]}=i\sin{[\theta]}$, we finally have the Green's
function for the infinite well potential
\begin{equation}
  G_{f,i}=
  \frac{2m}{\hbar^2 k \sin{[k L]}}
  \sin{[k(x_{f}-L)]\sin{[k x_{i}]}}.
\end{equation}

\bibliographystyle{model1a-num-names}

\end{document}